\documentclass[letterpaper, 10 pt, conference]{ieeeconf}  % Comment this line out if you need a4paper

\IEEEoverridecommandlockouts                              % This command is only needed if 
\usepackage{graphicx}
\usepackage{subfig}
\usepackage[english]{babel}
\usepackage[latin1]{inputenc}
\usepackage[]{amsmath}
\usepackage{amssymb}
\usepackage[]{bm}
\usepackage[]{booktabs}
\usepackage[]{multirow}
\usepackage{tikz}
\usepackage{mathrsfs} %for \mathscr{} command
\usepackage{epstopdf}
\usepackage{cancel}
\usepackage{enumerate}
\usepackage[electronic]{ifsym}
\usepackage{xcolor}

\newcommand{\f}[2]{\frac{#1}{#2}}
\newcommand{\smallmat}[1]{\ensuremath{\left[ \begin{smallmatrix}#1
    \end{smallmatrix} \right]}}

\newcommand{\sign}{\ensuremath{\mathrm{sign}}}

\newcommand\real{\ensuremath{{\mathbb R}}}
\newcommand\C{\ensuremath{{\mathcal C}}}
\newcommand\D{\ensuremath{{\mathcal D}}}

\newcommand\PiS{\ensuremath{{\Pi^{\star}}}}
\newcommand\SGN{\ensuremath{\overline{\textrm{sgn}}}}
\newcommand\A{\ensuremath{\mathcal{A}}}
\newcommand\Ba{\ensuremath{\mathcal{B}_{\mathcal{A}}}}

\newcommand\sPotEx{\ensuremath{\frac{1}{2}k\hat\theta^2}}

\newcommand\epsPhi{\ensuremath{\epsilon_\phi}}

\newtheorem{thm}{Theorem}
\newenvironment{theorem}{\begin{thm}\rm }{\end{thm}}
\newtheorem{remm}{Remark}
\newenvironment{remark}{\begin{remm}\rm }{\hfill \hspace*{1pt} \hfill $\lrcorner$\end{remm}}
\newtheorem{defi}{Definition}
\newenvironment{definition}{\begin{defi}\rm }{\end{defi}}
\newtheorem{lem}{Lemma}
\newenvironment{lemma}{\begin{lem}\rm }{\end{lem}}

\newtheorem{coroll}{Corollary}

\newtheorem{ass}{Assumption}
\newenvironment{assumption}{\begin{ass}\rm }{\end{ass}}
\newtheorem{propty}{Property}

\newtheorem{prop}{Lemma}

\newtheorem{fac}{Fact}

\newtheorem{cla}{Claim}

\graphicspath{{../img/},{/}}
\RequirePackage[normalem]{ulem} %DIF PREAMBLE
\title{\LARGE \bf
Global results on reset-induced periodic trajectories of planar systems
}

\author{Andrea Bisoffi\textsuperscript{1}, Fulvio Forni\textsuperscript{2}, Mauro Da Lio\textsuperscript{1} and Luca Zaccarian\textsuperscript{3}% <-this % stops a space
\thanks{Work supported in part by ANR under project LimICoS, contract number 12 BS03 005 01, by the iCODE institute, research project of the Idex Paris-Saclay, and by the University of Trento, grant OptHySYS. }%
\thanks{\textsuperscript{1} Dipartimento di Ingegneria Industriale, University of Trento, Italy
        {\tt\small \{andrea.bisoffi, mauro.dalio\}@unitn.it}}%   
\thanks{\textsuperscript{2} Department of Engineering, University of Cambridge, United Kingdom {\tt\small f.forni@eng.cam.ac.uk}}
\thanks{\textsuperscript{3} CNRS, LAAS, 7 avenue du Colonel Roche, F-31400 Toulouse, France and Universit\'{e} de Toulouse, 7 avenue du Colonel Roche, 31077 Toulouse cedex 4, France, and Dipartimento di Ingegneria Industriale, University of Trento, Italy
        {\tt\small zaccarian@laas.fr}}%
}

\begin{document}

\maketitle
\thispagestyle{empty}
\pagestyle{empty}

\begin{abstract}
We study the existence of asymptotically stable periodic trajectories induced
by reset feedback. The analysis is developed for a planar system.
Casting the problem into the hybrid setting, 
we show that a periodic orbit arises from the balance
between the energy dissipated
during flows and the energy restored by resets, at jumps.
The stability of the periodic
orbit is studied with hybrid Lyapunov tools. 
The satisfaction of the so-called hybrid basic conditions
ensures the robustness of the asymptotic stability.
Extensions of the approach to more
general mechanical systems are discussed.
\end{abstract}

\section{Introduction}
\label{sec:intro}

Starting from the important theorem of Poincar{\'e}-Bendixson,
many theoretical efforts have been made in the characterization of periodic orbits 
for planar continuous-time nonlinear systems,
motivated by the pervasive presence of oscillators
in electronics, mechanics and biology \cite{Hirsch1974,Grasman1987}.
A recent research direction seeks to extend this effort to the hybrid setting, 
namely to the context where, for a planar dynamical system, a suitable 
interplay of continuous flow and discrete jumps of the solutions leads to the existence of
attractive periodic hybrid trajectories. The relevance of this
topic in engineering is readily shown by the studies on 
bipedal robotic walking, 
where periodic hybrid trajectories arise from the 
combination of the free motion of the legs (continuous flow)
with the impulsive action of the impacts at ground contact
(discrete jumps)
\cite{Westervelt2007,teel2013stabilization}.

The paper provides a stability analysis of hybrid periodic trajectories for planar
mechanical systems based on the hybrid Lyapunov stability tools in \cite{goebel2012hybrid}.
The main motivation for the paper
comes from the literature on variable impedance actuators, typically adopted
in robotics. Strongly inspired by biological musculoskeletal systems,
these actuators have a tunable stiffness and/or damping,
which play a relevant role to improve motion efficiency
\cite{Gunther2013,lakatos2013nonlinear,lakatos2013modal,lakatos2014jumping}.
For multi-body systems with
frequency separation between first and subsequent natural modes,
\cite{lakatos2014switching} and \cite{lakatos2013nonlinear} show that periodic oscillations can be obtained by means 
of simple switching control laws tuned only on the first natural mode.
Taking advantage of a number of hybrid tools, 
we revisit and extend the results in \cite{lakatos2014switching}.
We model the dynamics in \cite{lakatos2014switching} as a hybrid
system and we show the existence of a
unique (hybrid) periodic orbit arising when the energy dissipated during flow balances the energy
restored by the reset-control action at a jump.
The stability analysis exploits hybrid Lyapunov methods.
In particular, the asymptotic stability of the periodic orbit
follows from the decay along system trajectories 
of a suitable Lyapunov function tailored on the
kinetic and potential energies just after and just before a jump.

The most relevant advantage of the approach 
is the intrinsic (in-the-small) robustness of asymptotic stability
\cite[Chapter~7]{goebel2012hybrid}, which
makes possible the use of the reset feedback law in applications.
The robustness of the design 
guarantees that the stability of the attractor persists, 
and is degraded with continuity, in the presence of small parameter perturbation
or when the instantaneous reset law is replaced by 
a (sufficiently) fast continuous actuation.

The asymptotic stability of the 
attractor holds for any parameters configuration that 
allows for a unique periodic orbit. This follows from the 
fact that the Lyapunov function is based on the mechanical energy
just after and just before a jump. 
Its minimum is represented on the phase space by the 
set of points such that the dissipated and restored mechanical energy are balanced.
Its decay is a natural consequence of
the mechanical features of the system. 
Indeed, no explicit characterization of the periodic orbit is required.

We anticipate that our Lyapunov-based analysis has similarities with the classical Poincar{\'e} analysis of periodic orbits. The level sets of the Lyapunov function are univocally identified by the points of the hyperplane at which resets occur. This hyperplane plays the role of a Poincar{\'e} section. 
Namely, along the portion of trajectory starting from and returning to this hyperplane, the overall decay of our Lyapunov function captures the convergence of the return map towards the fixed point.
The advantage of a Lyapunov analysis is the characterization of the basin of attraction of the periodic orbit. In this sense, our approach is close in nature to the analysis of the rimless wheel in \cite{saglamlyapunov}.

A promising future direction from our results is to exploit
the hybrid framework to provide
mixed continuous-discrete control strategies to
optimize motion efficiency.

The paper is organized as follows.
The hybrid dynamics is discussed in Section \ref{sec:sysDescr}.
Sections \ref{sec:PHT} and \ref{sec:stability} provide
conditions for the existence of periodic hybrid trajectories
and for their stability. Technical proofs are in the Appendix.
Simulations in Section \ref{sec:sims} illustrate
the convergence towards the unique hybrid periodic orbit
of the system. A comparison with the literature and further
discussions are reported in Section~\ref{sec:discConclFuture}. 

\section{System description}
\label{sec:sysDescr}

Based on \cite{lakatos2014switching}, consider the classical mass-spring-damper mechanical system
\begin{equation}
\label{eq:massSpringDamper}
m \ddot q + c \dot q + k (q - \theta) = 0
\end{equation}
with mass, damping and elastic constants respectively $m$, $c$, $k$.
$q$ is the displacement of the mass and $\theta$ is the control input. The elastic force provided
by the spring is proportional to the difference $q-\theta$.
The role of $\theta$ is to enforce a variation in
the stored potential energy of the spring.
Following~\cite{Gunther2013}, 
$\theta$ could model the effect
of the slow preloading of the spring during the flight
phase of a hopping robot, which is then released by a clutch mechanism when touching the ground.

In what follows $\theta$ is piecewise constant: it switches between $\theta \in \{-\hat{\theta}/2, \hat{\theta}/2\}$ 
when the trajectories of the system pass through the hyperplane defined by
$\{(q,\dot{q})\in \real^2\,|\, q-\theta = 0\}$.
$\hat{\theta}>0$ is a design parameter corresponding to 
the amount of potential energy loaded in the spring 
at switches. Switches on $\theta$ can be considered as the limit of
a very fast continuous action on the spring, a kick of energy, rapidly moving
$\theta$ from one value to the other in $\{-{\hat{\theta}}/{2}, \hat{\theta}/2 \}$.

With coordinates $x_1 := q-\theta$ and $x_2 := \dot{q}$, the 
dynamics of the system can be represented according to the hybrid formalism in~\cite{goebel2012hybrid} as follows. Since $\theta$ is constant, the flow dynamics reads
\begin{subequations}
\label{eq:hybsys}
\begin{equation}
\label{eq:flow}
\dot x= f(x) :=
\begin{bmatrix}
x_2 \\
-\f{c}{m} x_2 -\f{k}{m} x_1
\end{bmatrix},\,x\in\C.
\end{equation}
The flow set $\C$ enabling flow dynamics is given by
\begin{multline}
\label{eq:Cset}
\C=\{ (x_1,x_2) \in \real^2 \colon x_1 x_2 \le 0\} \\
\cup \{ (x_1,x_2) \in \real^2 \colon |x_1| \ge \hat \theta, x_1x_2 \ge 0\} \ .
\end{multline}
The jump dynamics reads
\begin{equation}
\label{eq:jump}
x^+ \in G(x) :=
\begin{bmatrix}
\hat \theta \,\SGN(x_2)\\
x_2
\end{bmatrix},\, x \in \D
\end{equation}
where
\begin{equation*}
\SGN(x_2)=
\begin{cases}
\text{sign}(x_2) & \  \text{ if } x_2\neq 0 \vspace{1mm}\\
\{1,-1\} & \  \text{ if } x_2=0 \ .
\end{cases}
\end{equation*}
The jump set $\D$ enabling jump dynamics is given by
\begin{equation}
\label{eq:Dset}
\D = \{ (x_1,x_2) \in \real^2 \colon  x_1= 0 \} \ .
\end{equation}
\end{subequations}

\begin{figure}[htbp]
\centering
\includegraphics[width=.8\linewidth]{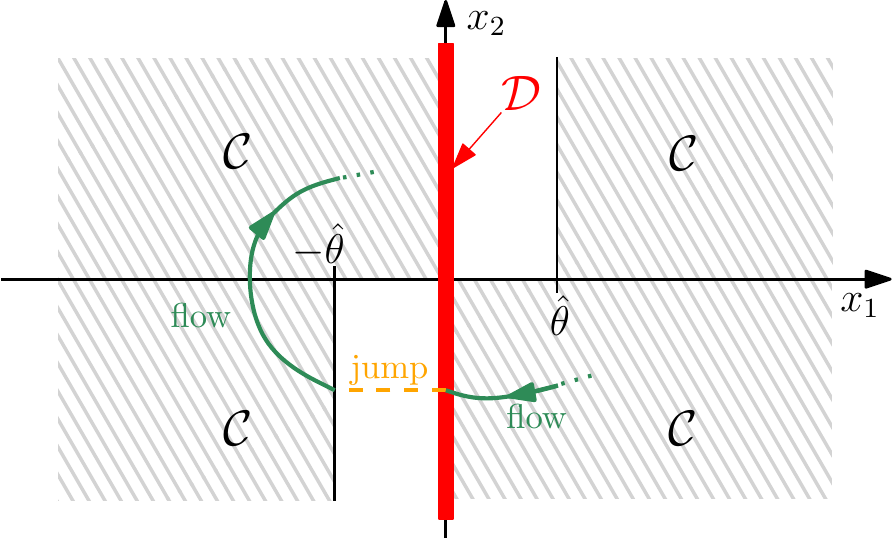}\vspace*{-1mm}
\caption{Flow set $\C$ and jump set $\D$ on the phase plane.}
\label{fig:CandD}
\end{figure} 

Figure \ref{fig:CandD} provides a graphical illustration of the
flow and jump set on the system phase plane. 
\eqref{eq:Dset}
guarantees that jumps occur when $q - \theta = x_1 = 0$. For $x_2 \neq 0$, we have that 
$|x_1|$ is reset from $0$ to $|x_1^+| = \hat{\theta}$ that is, 
$|q^+-{\theta}^+| = |q-{\theta}^+| = \hat{\theta}$.
Indeed, the reset corresponds to a switch in the equilibrium position of the spring, through actuation.
We do not reset the mass position $q$.

\begin{figure}[htbp]
\centering
\includegraphics[width=0.49\columnwidth]{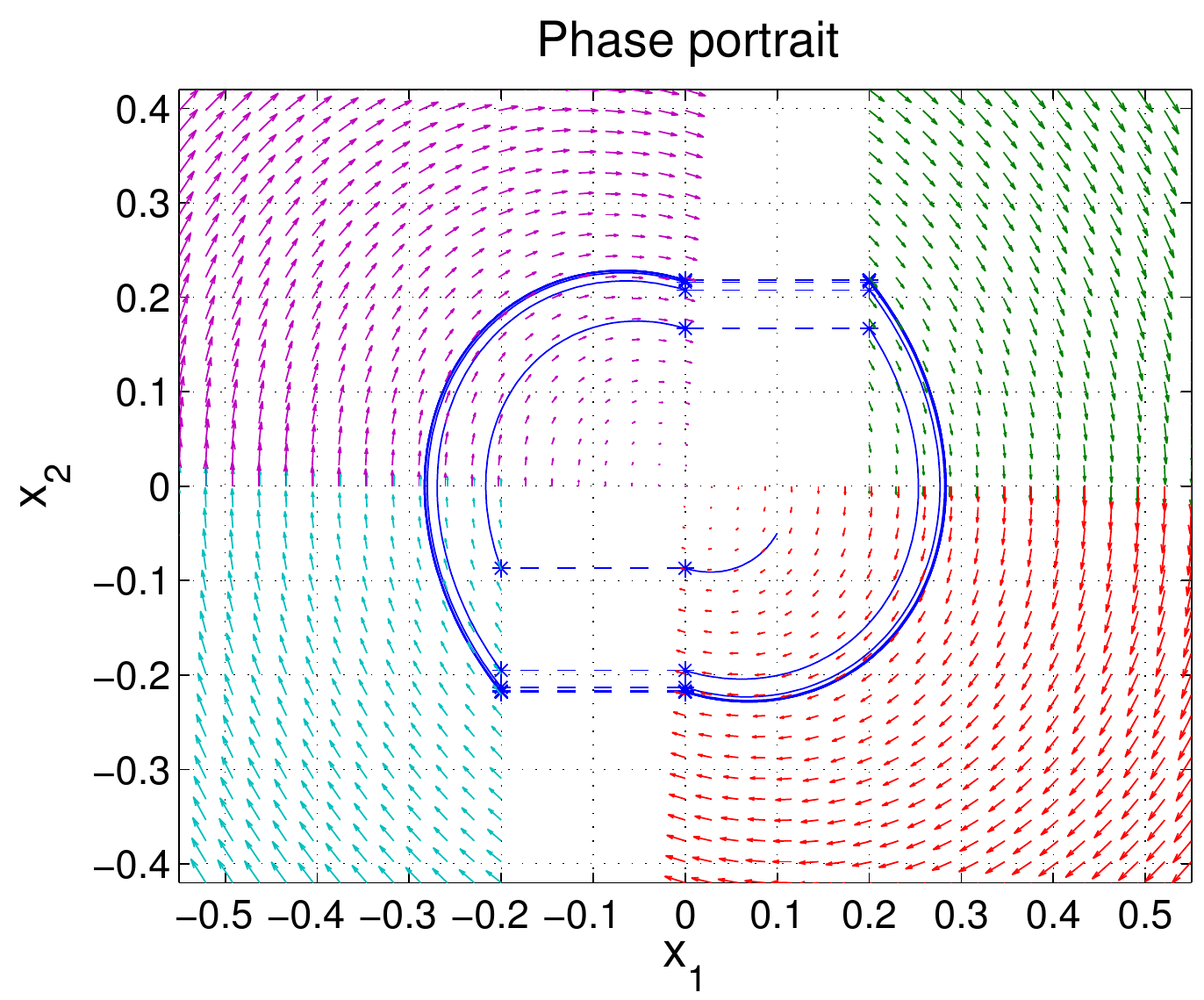} 
\includegraphics[width=0.49\columnwidth]{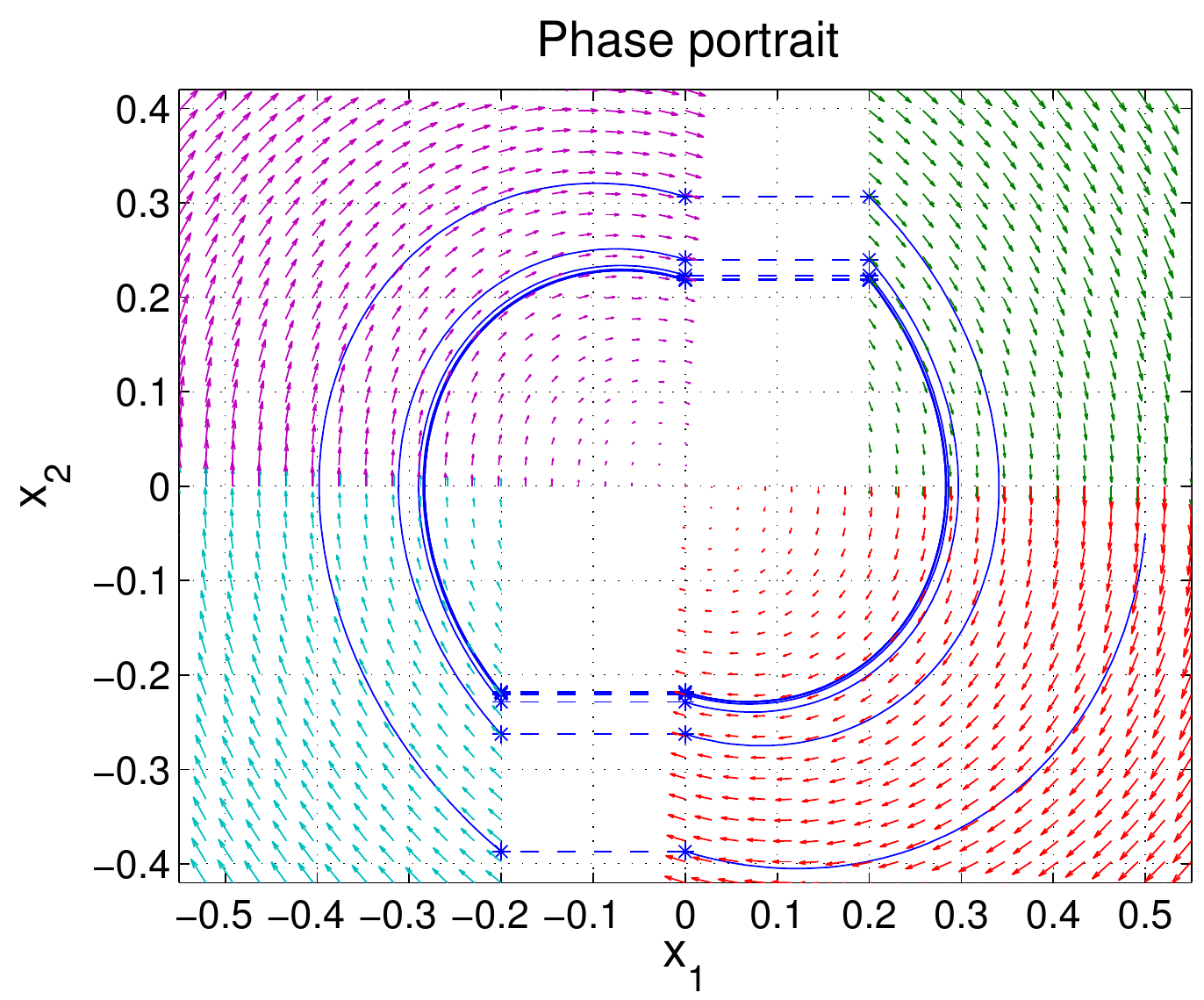} \vspace*{-1mm}
\caption{Phase plot of hybrid trajectories. Left: $(x_{1,0},x_{2,0}):=(0.1,-0.05)$. 
Right: $(x_{1,0},x_{2,0}):=(0.5,-0.05)$.
}
\label{fig:phasePlot}
\end{figure}

The behavior of the solutions is illustrated in Figure~\ref{fig:phasePlot}, which 
for a system with parameters
$m=1$ kg, $c=0.3$ Ns/m, $k=1$ N/m, $\hat \theta=0.2$ m, 
for two different initial conditions. The two trajectories
converge asymptotically to an attractor defined by the image
of a hybrid periodic trajectory, where periodicity must be 
intended in a hybrid sense as clarified in the next section.

\begin{remark}
In (\ref{eq:jump}) we used the set-valued mapping $\SGN$ to guarantee that the graph
of the jump map $x\mapsto G(x)$ is a closed set. This feature ensures the outer semicontinuity of $G$.
Outer semicontinuity of $G$ combined with the 
continuity of $f$ and with the fact that 
$\C$ and $\D$ are closed sets guarantees that hybrid system (\ref{eq:hybsys}) satisfies
the hybrid basic conditions \cite[Assumption 6.5]{goebel2012hybrid}.
They guarantee regularity of the solution set and robustness 
to small perturbations \cite{Goebel2006}.
\end{remark}

\section{Hybrid periodic orbits}
\label{sec:PHT}

The notion of periodicity for a hybrid trajectory is a straightforward extension 
of the usual notion of periodicity.
\begin{definition}
\label{def:PHT}
Given any hybrid system $\mathcal{H}:=(\C,F,\D,G)$, a \emph{hybrid periodic 
trajectory} is a complete solution $x$ for which there exists a pair 
$(T,J)$ with either $T\in\real_{\ge 0}$ and $J\in \mathbb{Z}_{>0}$ or $T\in\real_{> 0}$ and $J\in \mathbb{Z}_{\ge 0}$ 
such that $(t,j)\in \text{dom}(x)$ implies $(t+T,j+J)\in\text{dom}(x)$ and, moreover,
\begin{equation}
x(t,j) = x(t+T,j+J).
\end{equation}
The image of $x$ is a  \emph{hybrid periodic orbit}.
\end{definition}

The following standing assumption on the parameters 
of the hybrid dynamics (\ref{eq:hybsys}) is 
necessary for the existence of a
nontrivial\footnote{A nontrivial hybrid periodic orbit
comprises more than one point.} hybrid periodic trajectory.
\begin{assumption}
\label{assume:parameters}
$\hat \theta$, $m$, $c$ and $k$ are strictly positive.
The roots of $m s^2 + c s + k = 0$ are complex conjugate, that is
$\left( \f{c}{2m} \right)^2 - \f{k}{m}<0$.
\end{assumption}
\smallskip

Assumption~\ref{assume:parameters} guarantees that 
$m \ddot{x}_1+ c \dot{x}_1 +k x_1=0$ is an underdamped mechanical 
system \cite[Chap.~2.2]{meirovitchVibrations}. When the system is not underdamped 
there is no guarantee that a nontrivial hybrid periodic trajectory exists. 
With real eigenvalues in the flow map, 
the trajectories of the system may converge 
to the origin according to the direction of the eigenvector corresponding 
to the slowest eigenvalue, lying in the second/fourth quadrant for $m,c,k>0$. 
In such a case, the solutions to (\ref{eq:hybsys}) 
exhibit at most one jump and the origin is a globally asymptotically stable equilibrium.

\begin{figure}[htbp]
\centering
{\includegraphics[width=0.49\columnwidth]{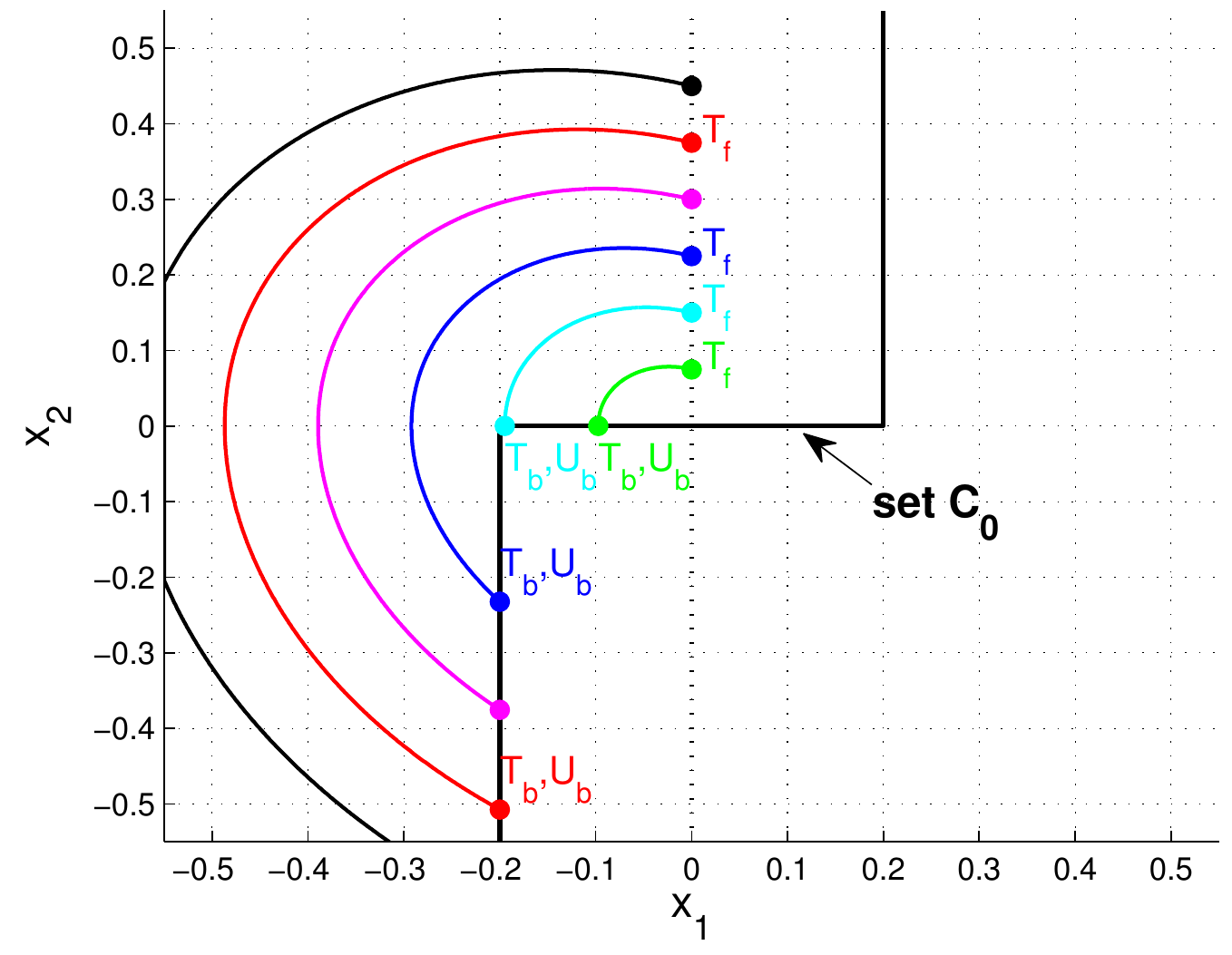}} 
{\includegraphics[width=0.49\columnwidth]{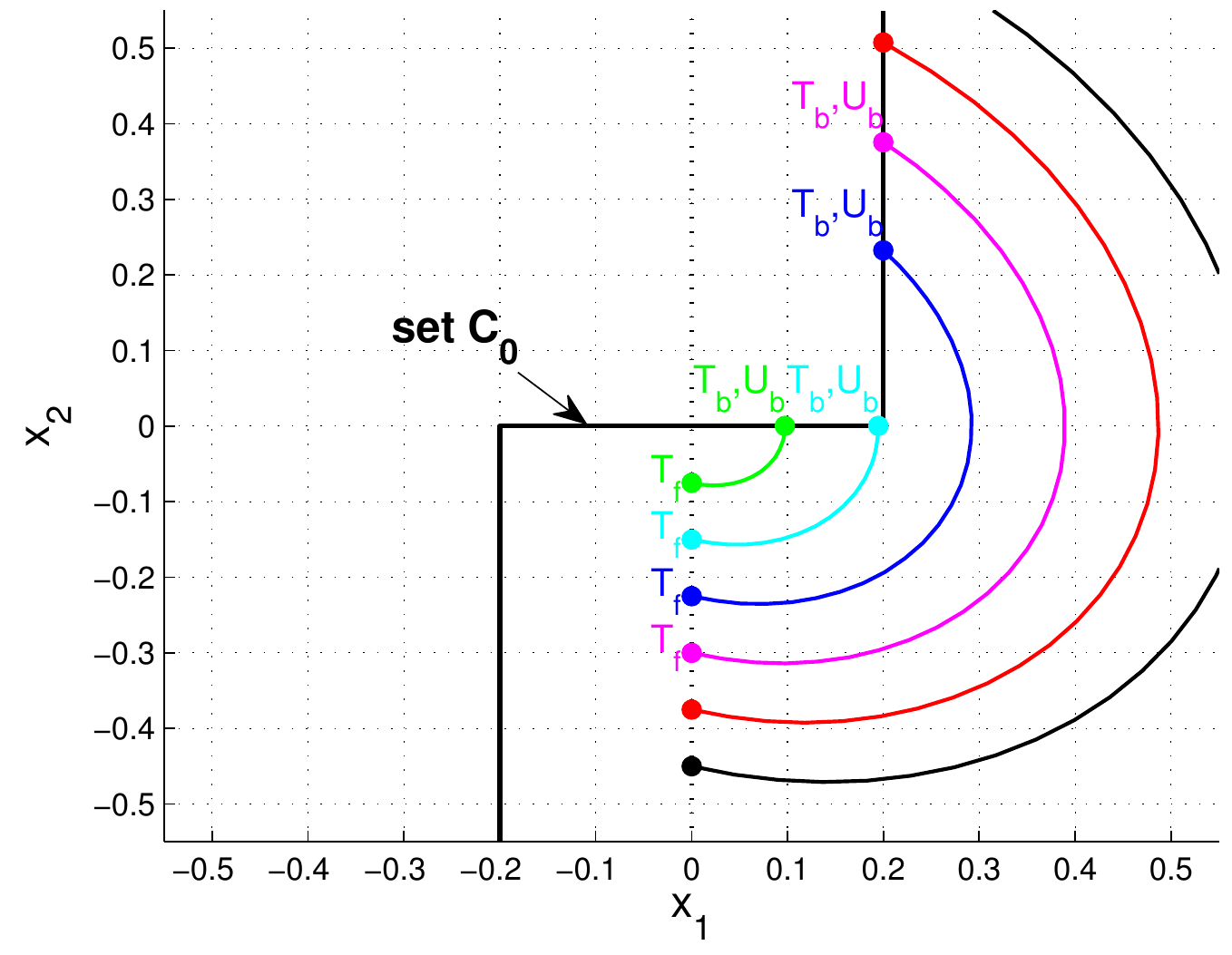}}
\caption{Set $\C_0$. Flowing solutions. The curves from $\C_0$
to $\D$ are level sets of $T_b$, $U_b$ and $T_f$.}
\label{fig:trajParam}
\end{figure}

The existence of a nontrivial hybrid periodic orbit follows
from energy considerations. Consider the $\rotatebox[origin=c]{90}{\FallingEdge}$-shaped
curve $\C_0$ represented in Figure \ref{fig:trajParam}, given by the set
\begin{multline}
\label{eq:C0def}
\C_0=\{x\in\real^2\colon |x_1|=\hat \theta, x_1x_2\ge 0\} \\
\cup \{x\in\real^2\colon |x_1| \le \hat \theta, x_2=0\}.
\end{multline}

Under Assumption \ref{assume:parameters},
the trajectories starting from $\C_0$ necessarily flow until
they reach $\D$. More specifically, 
flowing solutions from any $x\in \C$ 
in forward (respectively, backward) time reach set $\D$ (respectively, $\C_0$) in \emph{finite} time because of the revolving nature of the flow trajectories. The following quantities are thus well defined.
\begin{itemize}
\item \emph{Backward energies.} Denoting by $(x_{1b}, x_{2b})$ the intersection with $\C_0$ after flowing in backward time from $x$,\begin{subequations}
\label{eq:forwBackEnergies}
\begin{equation}
T_b(x) :=\f{1}{2} m x_{2b}^2 \quad\qquad U_b(x):=\f{1}{2} k x_{1b}^2
\end{equation}
are the \emph{backward kinetic} and \emph{backward potential} energies, respectively.
\item \emph{Forward energies.} Denoting by $(x_{1f}, x_{2f})$ the intersection with $\D$ after flowing in forward time from $x$, 
\begin{equation}
T_f(x):=\f{1}{2} m x_{2f}^2
\end{equation}
\end{subequations}
is the \emph{forward kinetic} energy.
\end{itemize}
Figure \ref{fig:trajParam} shows the level sets of $T_b$, $U_b$ and $T_f$, which correspond indeed to flowing portions of solutions to (\ref{eq:hybsys}).

For each $x\neq 0$ the quantity $T_b(x)+U_b(x)$ is the total mechanical energy of the system
right after a jump. The quantity $T_f(x)$ is the total mechanical energy of 
the system after a maximal\footnote{In the same sense of \emph{maximal} solutions in \cite[Definition~2.7]{goebel2012hybrid}, namely that it can not be extended further.} flow, that is, right before a jump.
The difference between these two energies corresponds to the 
dissipation during flows. 

The reset of $\theta$ injects
energy into the system in the form of potential energy. This fact and the central symmetry of the phase portrait (namely, if $\phi$ is a solution to (\ref{eq:hybsys}), $-\phi$ is a solution as well) imply that a hybrid periodic orbit corresponds to the set of points satisfying the energy balance
\begin{equation}
\label{eq:energy_balance}
T_b(x) + U_b(x) = T_f(x) + \frac{1}{2} k \hat{\theta}^2 \,
\end{equation}
where the last term represents precisely the potential energy injected by
a reset. Given the mentioned central symmetry, $x\neq 0$ belongs to a \emph{periodic} hybrid orbit only if $U_b(x) = \sPotEx$, so that (\ref{eq:energy_balance}) is equivalent to
$T_b(x) = T_f(x)$.

Existence and uniqueness 
of the hybrid periodic orbit follows from the 
monotonicity of the dissipation with respect to initial
conditions on $\C_0$. In fact, for $x=(x_1,x_2) \in \C_0$
the dissipation is a strictly increasing function of $|x|$. This is precisely stated in the next lemma, proven in the Appendix.

\begin{lemma}
\label{lem:dissArea}
Consider any solution $x$ to (\ref{eq:hybsys})
flowing from $\C_0$ at ordinary time $t_1$ to $\D$ at ordinary time $t_2\geq t_1$ and define
the total mechanical energy at $x$ as $E(x)=\frac{1}{2} m x_2^2 + \frac{1}{2} k x_1^2$. 
The dissipated energy $E(x(t_1))-E(x(t_2))$ is equal to $c \Pi$, where $\Pi$ is the (unsigned) area within the curves given by 
the image of the solution, the set $\C_0$ and the set $\D$
(hatched area in Figure \ref{fig:dissArea}).
\end{lemma}

The monotonicity of the dissipation is clear from Figure \ref{fig:dissArea}.  
As a consequence, there is only one initial condition on $\C_0$ 
for which \eqref{eq:energy_balance} holds.
\begin{theorem}
\label{thm:existence_uniqueness}
Under Assumption \ref{assume:parameters}, there exists
a unique nontrivial hybrid periodic orbit for the hybrid system \eqref{eq:hybsys}.
$T_b(x) = T_f(x)$ at each point $x$ of the hybrid periodic orbit.

\begin{figure}[htbp]
\centering
\includegraphics[width=.8\linewidth]{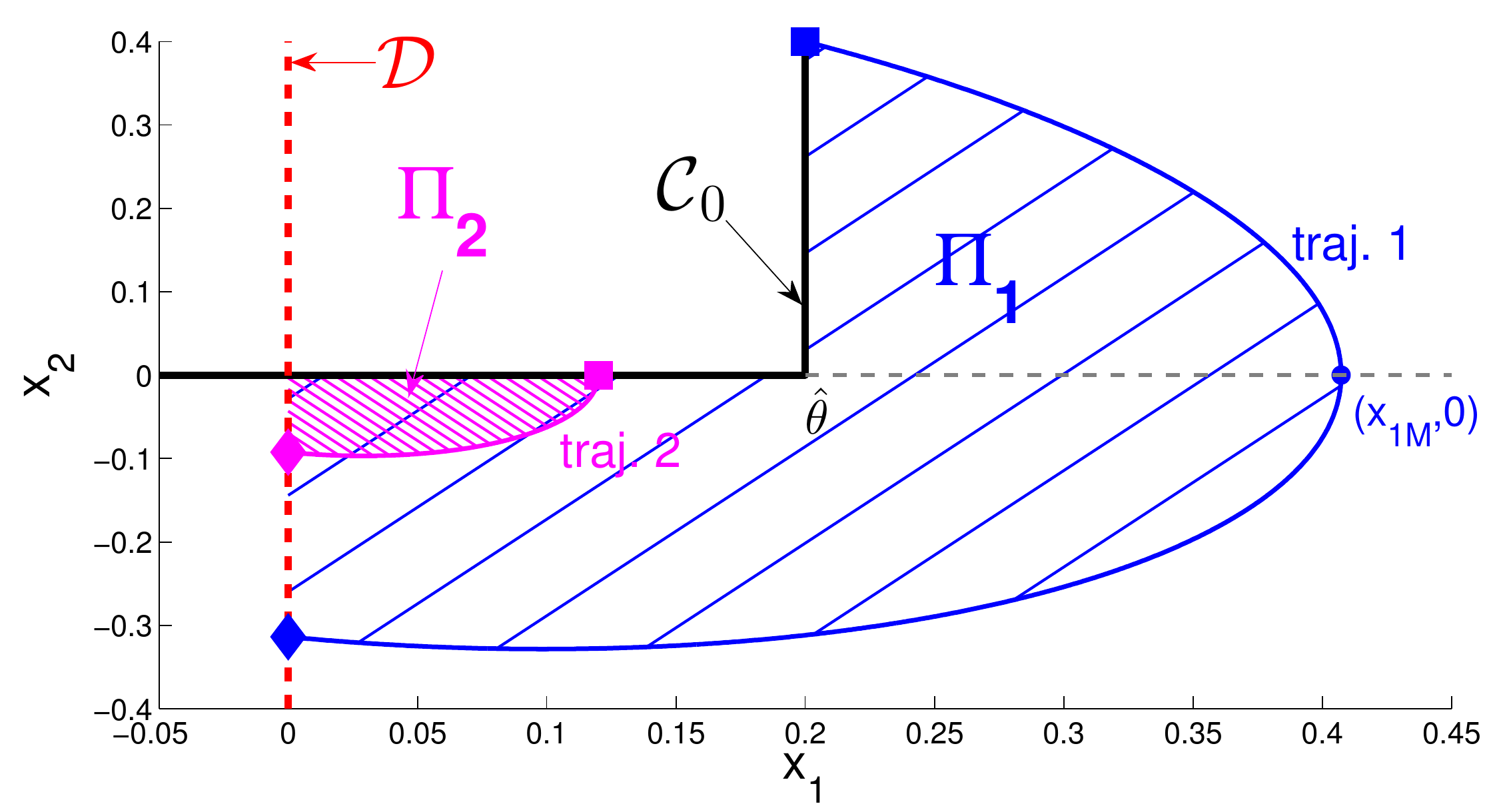}
\caption{The hatched areas is proportional to the energy dissipated by damping.}
\label{fig:dissArea}
\end{figure}
\end{theorem}

\section{Global asymptotic stability}
\label{sec:stability}

The stability of the nontrivial hybrid periodic orbit 
is a set stability problem.
Consider the attractor given by 
\begin{equation}
\label{eq:setA}
\A=\{x\in\C \colon T_b(x)=T_f(x), x \neq 0\} \ .
\end{equation}
Energy considerations similar to those in the previous section
readily show that $\A$ is compact and forward invariant (see the proof of Lemma~\ref{lem:V}). The images of all nontrivial hybrid periodic trajectories of (\ref{eq:hybsys}) coincide with $\A$.
Convergence and stability of the periodic motion follow from
the next theorem.
\begin{theorem}
\label{thm:main}
Under Assumption \ref{assume:parameters}, the set $\A$ in (\ref{eq:setA}) 
is an asymptotically stable attractor for the hybrid system \eqref{eq:hybsys}
with basin of attraction $\Ba=\real ^2 \backslash \{0\}$. 
\end{theorem}

The origin $x=0$ is not in $\Ba$ because it is a weak equilibrium:
solutions to (\ref{eq:hybsys}) starting from the origin can flow forever staying at the origin or may jump to $-\hat \theta$ or $\hat \theta$ and then converge to the hybrid periodic orbit. 

We remark that 
the stability of the set $\A$ does not require an
explicit characterization of the hybrid periodic orbit. 
We only need to ensure the feasibility of the 
balance in \eqref{eq:energy_balance}. Therefore, by
Theorem \ref{thm:existence_uniqueness} and
Theorem \ref{thm:main},
the reset feedback law induces a 
hybrid periodic trajectory for every
parameters selection that satisfies
Assumption \ref{assume:parameters}. Then future work comprises performing optimal selections of $\theta$ (and the arising periodic motion) via hybrid adaptation.

\begin{figure}[htbp]
\centering
{\includegraphics[width=0.9\columnwidth]{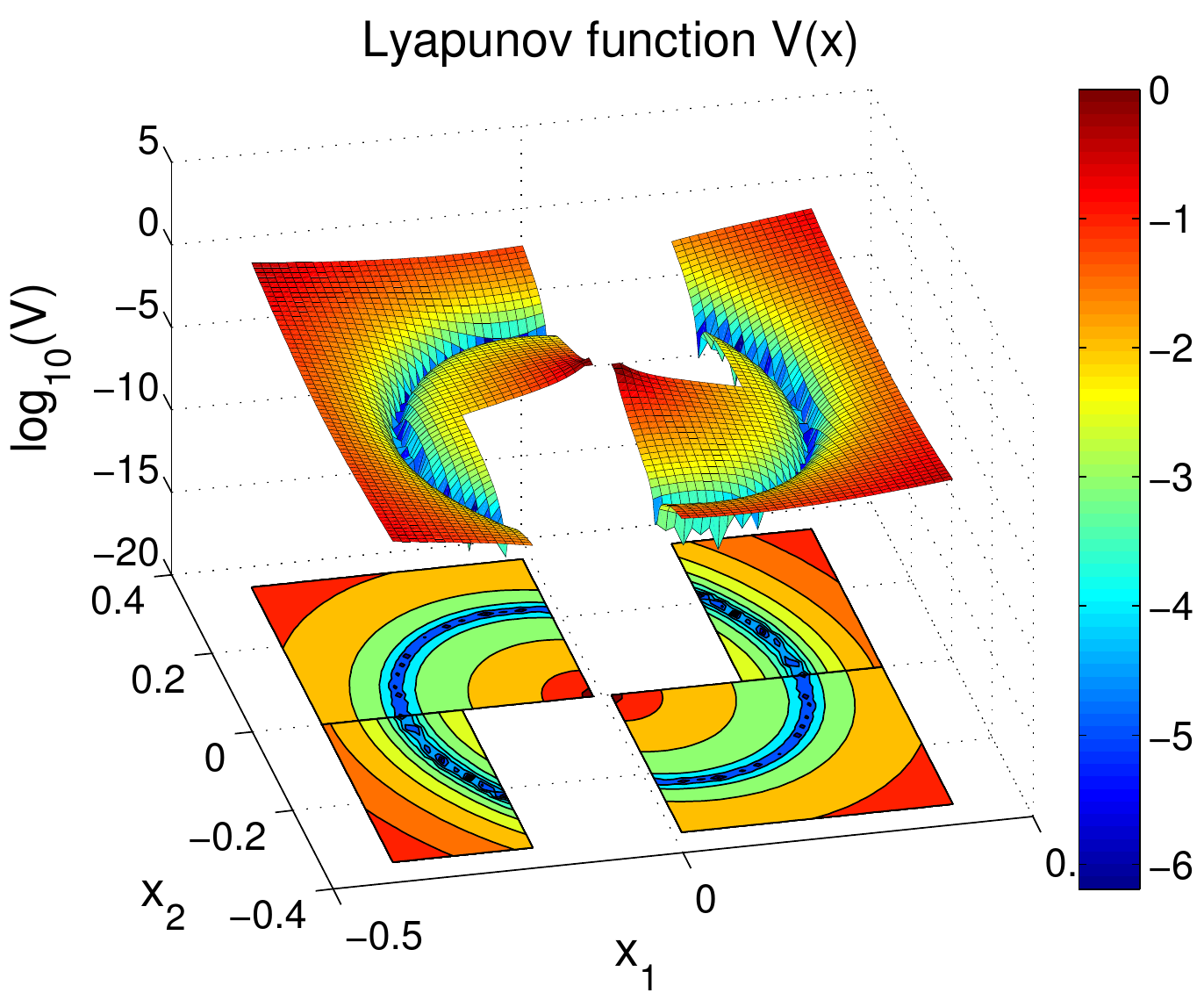}}
\caption{The Lyapunov function $V$ in logarithmic scale.}
\label{fig:VofX}
\end{figure}

The proof of Theorem \ref{thm:main} is based on a Lyapunov argument.
Using the definitions in \eqref{eq:forwBackEnergies}, 
consider the Lyapunov function candidate for $\A$
given by
\begin{equation}
V(x)=\f{(U_b(x) + T_b(x) - T_f(x)-\sPotEx)^2}{U_b(x)} \ .
\label{eq:VofX}
\end{equation}
The shape and the level sets of $V$ are illustrated in Figure \ref{fig:VofX}.
Note that the Lyapunov function $V$ blows up as $x$ approaches $0$ (the boundary of $\Ba$) 
and as $x$ grows unbounded, as shown in Figure~\ref{fig:VofX} for the same parameter
selection of Figure~\ref{fig:phasePlot}.
\vspace*{-2mm}
\vspace*{-2mm}

The next lemma is a key step for proving Theorem~\ref{thm:main}.
\begin{lemma}
\label{lem:V}
Under Assumption~\ref{assume:parameters}, the 
set $\A$ in (\ref{eq:setA}) is nonempty and compact and the Lyapunov function $V$ in (\ref{eq:VofX})
\begin{subequations}
\begin{enumerate}[(i)]
\item is positive definite with respect to $\A$ on $\Ba\cap(\C\cup\D)$, namely \label{lem:V1}
\begin{equation}
\label{eq:VdefPos}
\begin{aligned}
V(x) & =0 \text{ if } x\in\A\\
V(x) & >0 \text{ if } x\in \Ba\cap(\C\cup\D)\backslash\A\\
\lim_{\substack{|x |\rightarrow 0^+\\ |x|\rightarrow +\infty}} V(x) & =+\infty
\end{aligned}
\end{equation}
\item is constant in the flow direction\footnote{We do not give a formal proof of smoothness of $V$ in our derivation, therefore it would be more appropriate to use the directional derivative of $V$ in (\ref{eq:VconstFlow}). We use here the notation with the gradient to keep the discussion simple.} \label{lem:V2}
\begin{equation}
\left< \nabla V(x), f(x) \right>=0,\,\, \forall x \in\C
\label{eq:VconstFlow}
\end{equation}
\item provides strict decrease across jumps\footnote{Note that $\SGN(x_2)$ is single-valued for all $x\in\D\cap\Ba=\{x\in\real^2\colon x_1=0, x_2\neq 0\}$ because it is set-valued only at $x_2=0$.} \label{lem:V3}
\begin{equation}
\label{eq:VdecreseFlow}
V\left(G(x)\right) - V(x) < 0,\,\,\forall x\in\Ba\cap\D\backslash\A.
\end{equation}
\end{enumerate}
\end{subequations}
\end{lemma}

\medskip

\begin{remark}
Following \cite[Corollary 7.32]{goebel2012hybrid} and \cite[Definition 7.29]{goebel2012hybrid}), 
item (\ref{lem:V1}) of Lemma~\ref{lem:V} implies that for any indicator function 
$\omega$ of $\A$ on $\Ba$ there exists class $\mathcal{K}_\infty$ functions 
$\underline{\alpha}$ and $\overline{\alpha}$ such that 
\begin{equation}
\label{eq:Vsandwich}
\underline{\alpha}(\omega(x))\le V(x) \le \overline{\alpha}(\omega (x)),
\end{equation}
which entails standard additional features of $\A$, e.g., robustness and semiglobal-practical robustness of asymptotic stability of $\A$
\cite[Chap.~6.7]{goebel2012hybrid}. 
\end{remark}

Based on Lemma~\ref{lem:V}, the proof of Theorem~\ref{thm:main} is a mere application of fundamental Lyapunov results holding for hybrid systems described with the formalism in \cite{goebel2012hybrid}. In particular, Lemma~\ref{lem:V} establishes that function $V$ in (\ref{eq:VofX}) is a non-strict Lyapunov function for compact attractor ${\mathcal A}$, in the sense of \cite[Equations (6a), (7)]{PrieurTAC14}. 
Indeed, the fact that ${\mathcal A}$ is compact is sufficient to obtain that (\ref{eq:VdefPos}) implies
(\ref{eq:Vsandwich}) for any indicator function $\omega$ and suitable class ${\mathcal K}_{\infty}$ functions $\underline \alpha$, $\overline \alpha$. Moreover, (\ref{eq:VconstFlow}) coincides with \cite[Equation (7a)]{PrieurTAC14} and (\ref{eq:VdecreseFlow}) implies \cite[Equation (7b)]{PrieurTAC14} for a suitable positive definite function $\rho$, once again because ${\mathcal A}$ is compact.
As a consequence, we may apply a local generalization of \cite[Theorem 2]{PrieurTAC14} by noticing that all solutions to (\ref{eq:hybsys}) are semiglobally persistently jumping, namely for any arbitrarily large number $\Delta$, we have that solutions restricted to flow\footnote{As customary, we denote by ${\mathbb B}$ the closed unit ball.} in ${\cal C}\cap{\Delta {\mathbb B}} \cap {\mathcal B}_{\mathcal A}$ and jump from ${\cal D}\cap{\Delta {\mathbb B}} \cap {\mathcal B}_{\mathcal A}$ have a uniform reverse dwell time (i.e., a maximum time between each pair of consecutive jumps in their domain). Such a maximum time is easily computed as the minimum flow time from the vertical portion of ${\mathcal C}_0$ intersected with $\Delta {\mathbb B}$. Note, in particular, that by homogeneity of the linear flow equation, such a maximum flow time is strictly smaller than the flow time of the first flowing interval of the (unique) solution starting from $x(0) = \smallmat{\hat \theta\\ \Delta}$.

\section{Simulations}
\label{sec:sims}

\begin{figure}[htbp]
\centering
\includegraphics[width=0.49\columnwidth]{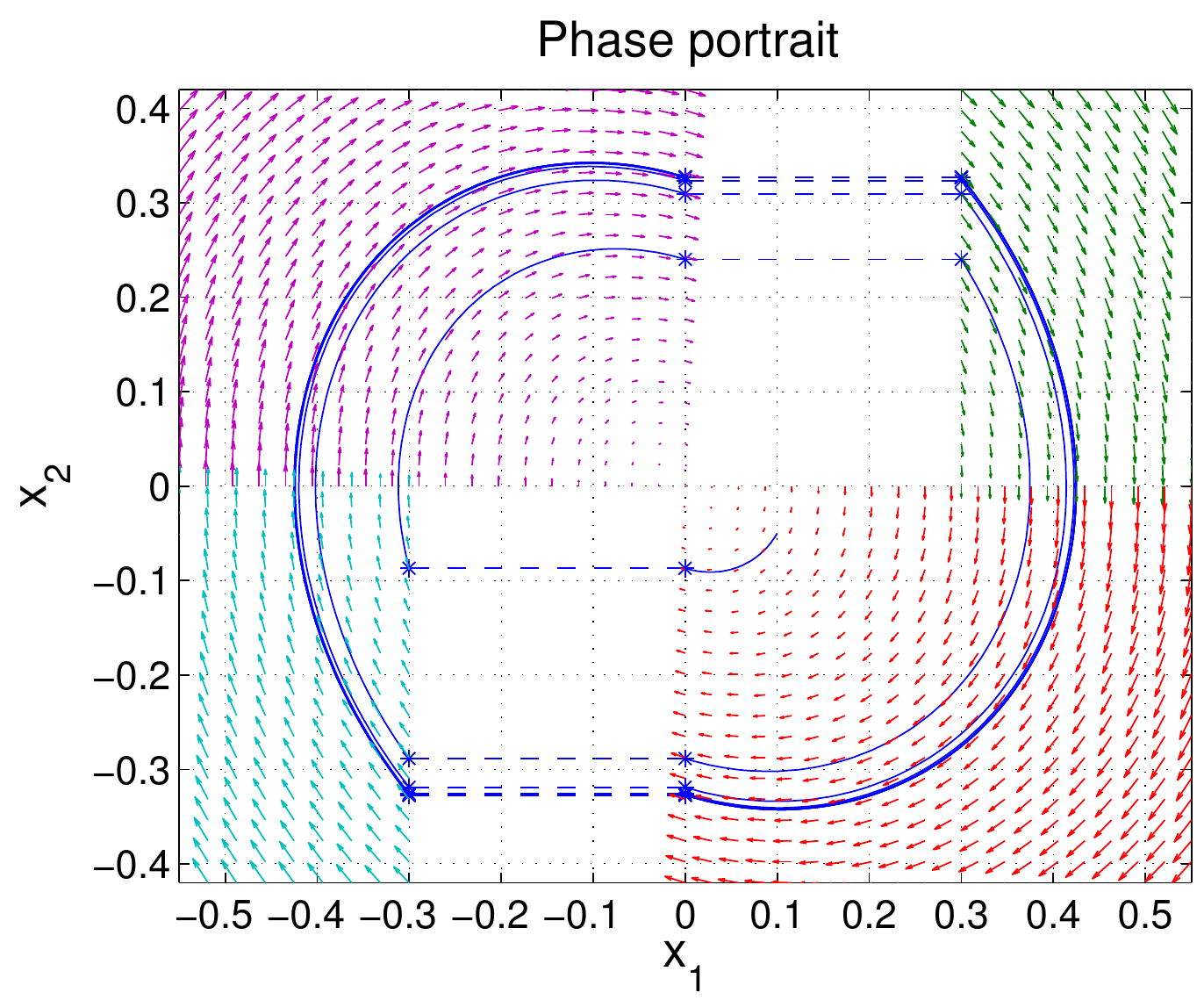}
\includegraphics[width=0.49\columnwidth]{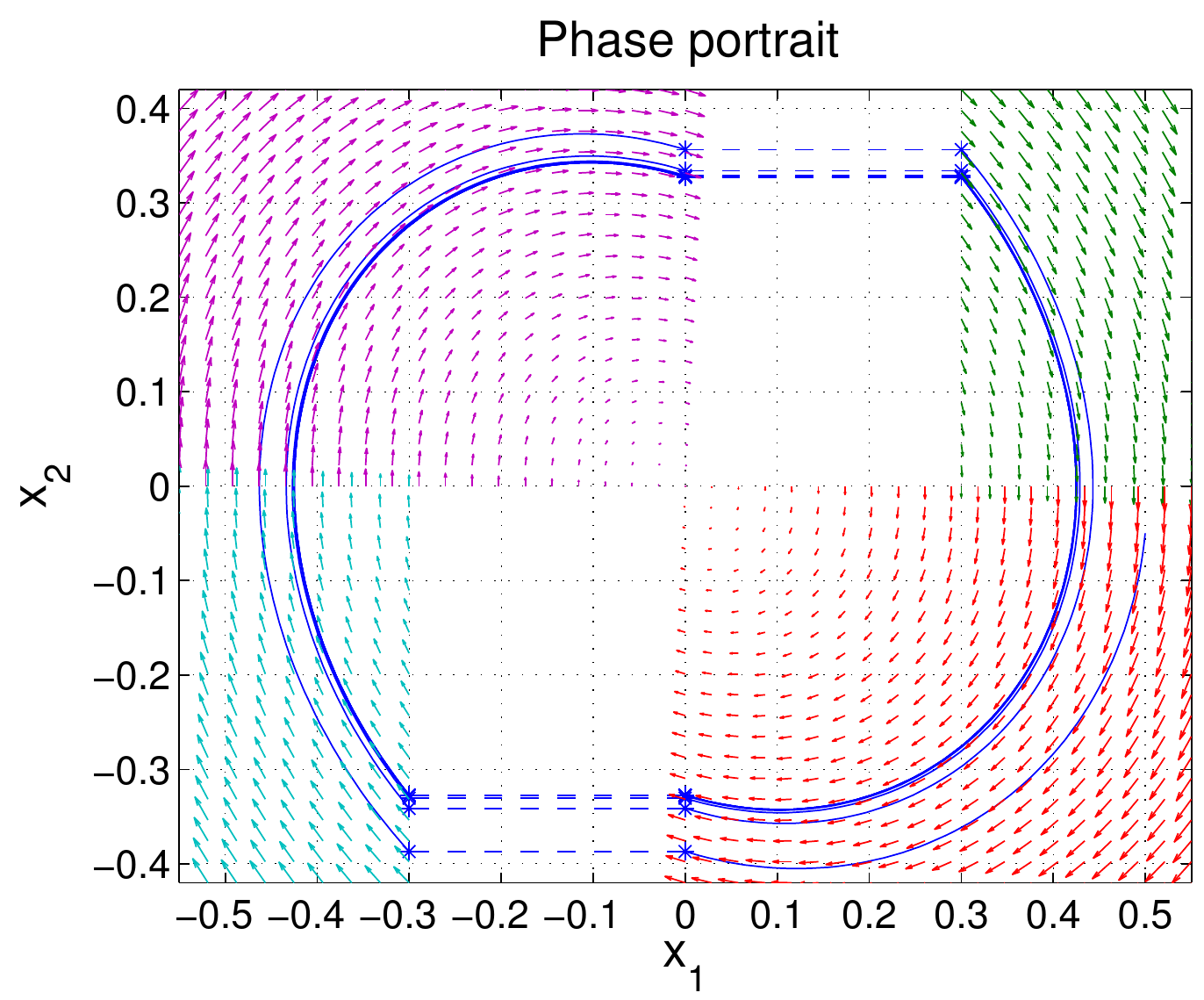}
\includegraphics[width=0.49\columnwidth]{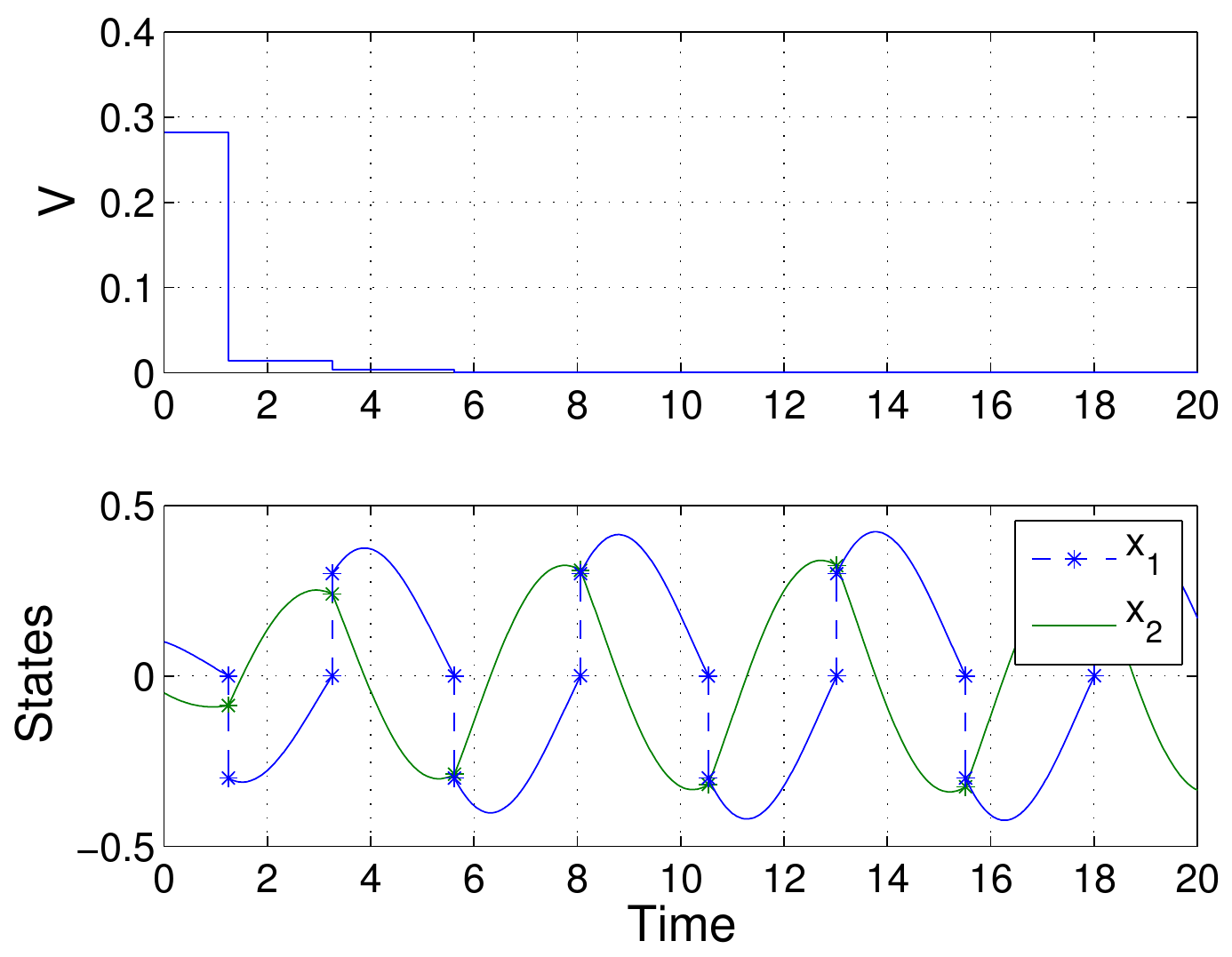}
\includegraphics[width=0.49\columnwidth]{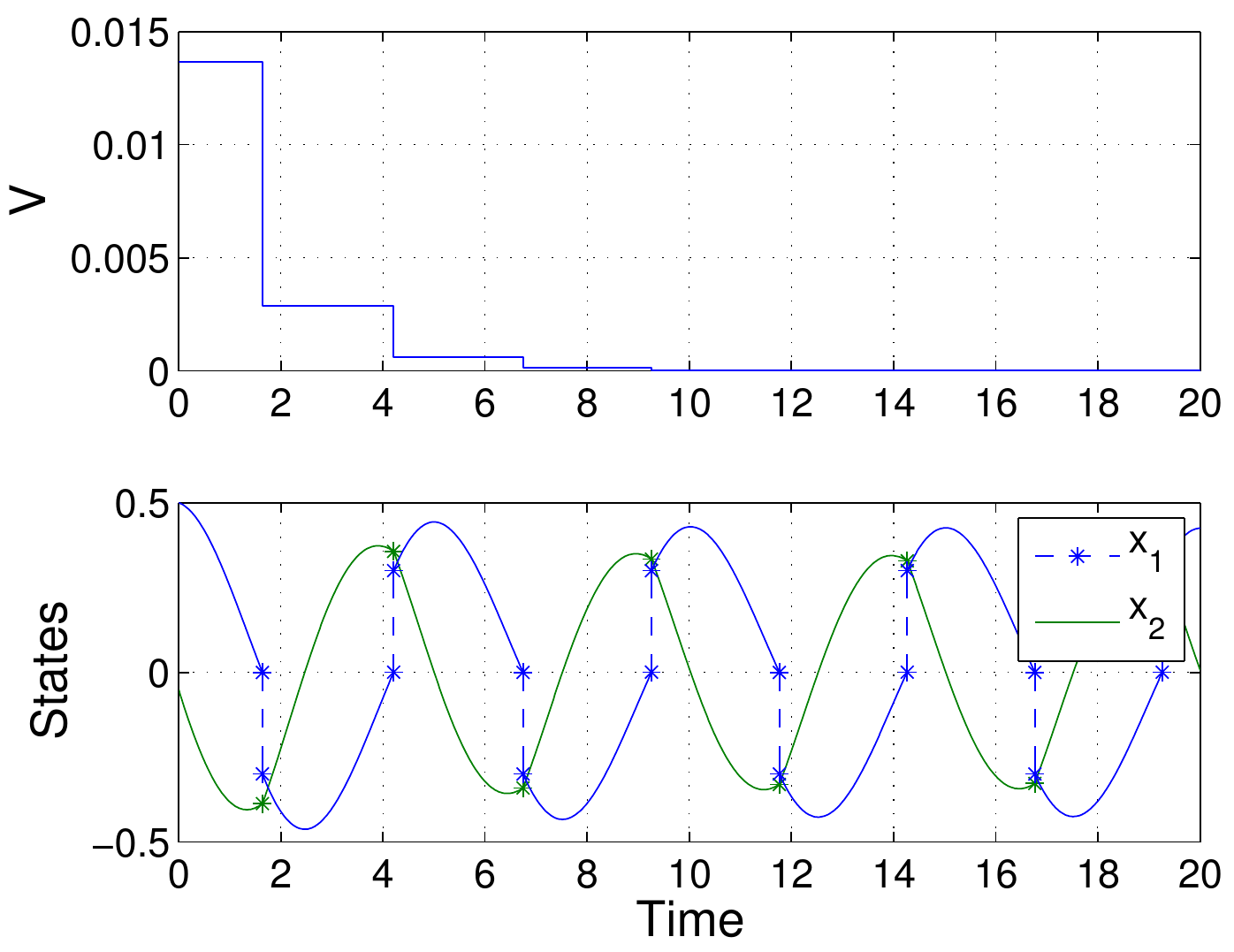} 
\caption{Upper part: phase plot of the trajectories.
Bottom part: Lyapunov function and hybrid trajectories projected on ordinary time for the two initial conditions.
Left: initial condition $(0.1,-0.05)$.
Right: initial condition $(0.5,-0.05)$.
}
\label{fig:simulations} 
\end{figure}

We simulate the hybrid system \eqref{eq:hybsys} 
with the same parameters adopted in Figure \ref{fig:phasePlot}, 
corresponding to $m=1$~kg, $c=0.3$~Ns/m, $k=1$~N/m (eigenvalues $s_{1,2}=-0.15 \pm i 0.9887$, 
consistent with Assumption~\ref{assume:parameters}).
Instead of $\hat \theta=0.2$~m we enforce a reset to the
larger value $\hat \theta=0.3$. Compared to Figure \ref{fig:phasePlot},
the hybrid periodic orbit changes as shown in the upper part of 
Figure \ref{fig:simulations}.
The bottom part of Figure \ref{fig:simulations} shows the values of the Lyapunov function and of the states $(x_1,x_2)$ as functions of ordinary time. 

\section{Conclusions and future research}
\label{sec:discConclFuture}

\begin{figure}[htbp]
\centering
\subfloat[][Flow and jump sets. \label{fig:CandDdlrII}]
{\includegraphics[width=0.45\columnwidth]{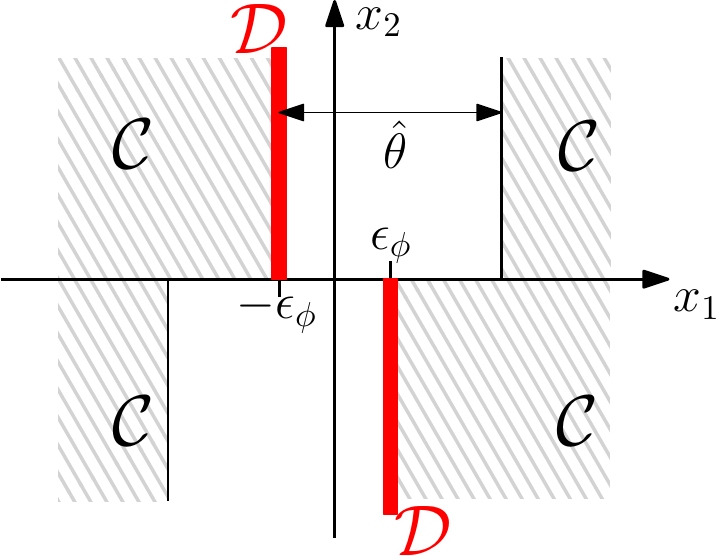}} \,
\subfloat[][Trajectories. \label{fig:trajdlrII}]
{\includegraphics[width=0.53\columnwidth]{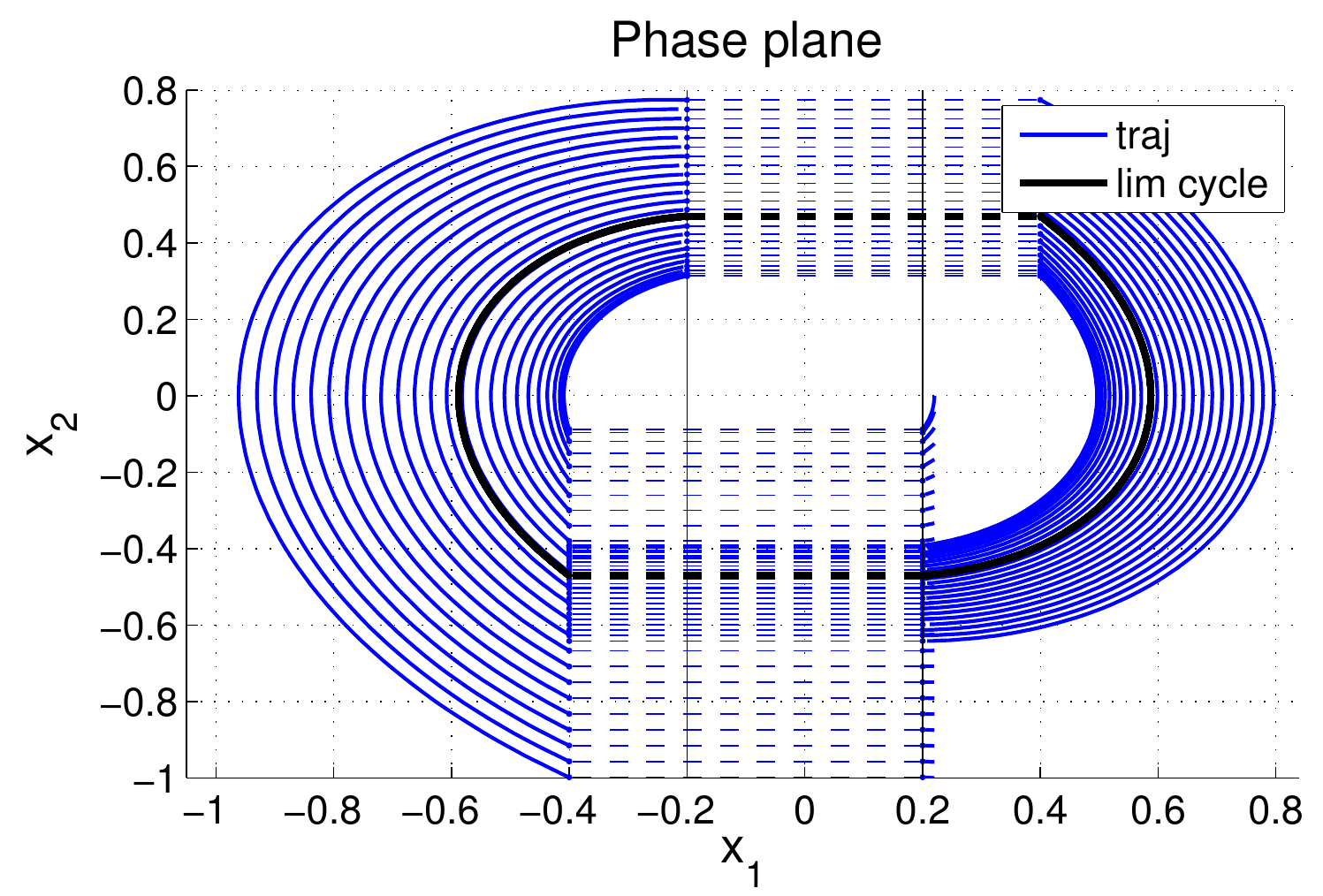}} \\
\caption{$\epsilon_\phi = 0.2$ and $\hat \theta=0.6$.}
\label{fig:case1jumpMGC} 
\end{figure}

Following \cite{lakatos2014switching},
different hybrid periodic orbits can be generated by 
exploiting different reset laws. 
Within the hybrid characterization (\ref{eq:hybsys}), 
one of the switching laws proposed in \cite{lakatos2014switching} 
is captured by the flow and jump sets $\C$ and $\D$ 
given in Figure~\ref{fig:CandDdlrII}: $\hat \theta \ge 2 \epsPhi$ and 
$\epsPhi>0$  is a new control parameter.
$G(x)=\!\smallmat{x_1+\hat \theta \sign(x_2)  \\ x_2}$ and
$f(x)$ remains \eqref{eq:flow} with the parameters of Section~\ref{sec:sims}.
The simulation of a number of trajectories is reported in Figure~\ref{fig:trajdlrII}.
The approach and results in Sections \ref{sec:PHT} and \ref{sec:stability}
can be extended to capture the stability properties of this new hybrid system,
to show the existence of a globally asymptotically stable
hybrid periodic orbit.
Figure \ref{fig:case1jumpMGC}
shows the potential of the hybrid characterization 
in \eqref{eq:hybsys}. A number of reset feedback laws can be 
modeled by variations on the definitions of the sets $\C$ and $\D$ and of the jump map $G$. 
The monotonicity of the dissipated
energy is crucial for the uniqueness of
the attractor. Future research will investigate the 
family of reset feedback that guarantees existence and 
uniqueness of an asymptotically stable hybrid periodic orbit.

It is natural to consider mechanical systems with elastic potentials typical of nonlinear springs, so that \eqref{eq:flow} would read $\dot{x}_2=-\frac{c}{m}x_2 - \frac{1}{m}\frac{\partial U}{\partial x_1}$ where the potential $U$ is any positive definite function.
A necessary condition for the existence of an attractor with basin $\real^2\backslash \{ 0 \}$
is the strict monotonicity of the elastic potential, namely 
the positivity of the function $x_1 \mapsto x_1 \f{\partial U(x_1)}{\partial x_1}$. 
In fact, the lack of strict monotonicity may lead to the coexistence of
several attractors. 
The minimal requirement for the existence of globally attractive
hybrid periodic orbits is that the linearization 
of the flow dynamics at the origin has complex conjugate eigenvalues, which is a straightforward generalization of Assumption~\ref{assume:parameters}.
Otherwise, trajectories that are sufficiently close to the origin would flow towards 
it remaining in the second/fourth quadrant, without triggering any reset. 
Future work will look for tight sufficient conditions for the existence of a global 
attractor for general elastic potentials.

Finally, the generality of a Lyapunov approach for stability
analysis calls for extensions of the method to 
more general mechanical systems. The first step is the
analysis of $n$-dimensional linear mechanical systems perturbed by resets. 
We also recall that the model in \eqref{eq:hybsys},
albeit an abstraction, stems from the robotics context.
Further detailed analysis will investigate its potential for robotic applications, 
in particular the one of legged locomotion.

\appendix
\subsection{Proof of Lemma~\ref{lem:dissArea}}

The work performed by the nonconservative viscous force $F_d=c x_2$ in moving the point mass from position 1 to position 2 causes a change $E_2-E_1$ in the total mechanical energy \cite[Page~9]{meirovitchVibrations}. With the aid of Figure~\ref{fig:dissArea}, let us exemplify it on trajectories like 1 and denote $x_{1M}$ the value at which a trajectory crosses the line $x_2=0$,  so that \emph{on the trajectory} $x_2$ can be expressed as function of $x_1$ for each half-plane $x_2>0$, $x_2<0$. Then, by splitting the integral relative to the work in two pieces, we get $E(x(t_2))-E(x(t_1))=\int^{x_{1M}}_{\hat \theta} \!\! -(c x_2) d x_1 + \int_{x_{1M}}^0 \!\!\! -(c x_2) d x_1 = - c \Pi$.

\subsection{Proof of Lemma~\ref{lem:V}}
\label{sec:proofV}

The proof of this fact relies on Lemma \ref{lem:dissArea}. 
From (\ref{eq:VofX}),
\begin{align}
\nonumber
V(x) &= \frac{\Big(\overbrace{U_b(x)+T_b(x)}^{E_b(x)} - 
\overbrace{T_f(x)}^{E_f(x)} - \overbrace{\sPotEx}^{\hat{U}}\Big)^2}{U_b(x)}\\
     &= \frac{(c\Pi(x) - \hat{U})^2}{U_b(x)},
\label{eq:newV}
\end{align}
where 
$E_b$ and $E_f$ are the total energies of the system right after and right before a jump, respectively; 
$\hat{U}$ is the potential energy at
$\smallmat{\hat \theta\\ 0} \in \C_0$;
$\Pi(x)$ is the area spanned by the solution passing through $x$ during a
flow from $\C_0$ to $\D$ (where $E_f$ is evaluated);
and $c>0$ is the damping coefficient in \eqref{eq:massSpringDamper}.
Figure~\ref{fig:dissArea} provides two examples for $\Pi(x)$.

We are now ready to prove Lemma \ref{lem:V}.
First notice that, due to the uniqueness of flowing solutions, the function
$x\mapsto \Pi(x)$ is necessarily strictly increasing as $x$ moves farther from the origin (or any compact set). Denote by $c \Pi_0 = c \Pi\left( \smallmat{\hat \theta \\ 0}\right)$ the dissipated energy when starting from the corner of set $\C_0$ in (\ref{eq:C0def}). Moreover, denote by $c \PiS = \sPotEx = \hat{U}$ the dissipated energy associated to the hybrid periodic orbit.
Note that $\PiS > \Pi_0$ necessarily, because $c \Pi_0$ cannot be larger than the total energy $\hat{U}=c\PiS$ at the beginning of the corresponding solution starting from the corner of $\C_0$.
Then,
\begin{equation}
\label{eq:setAequiv}
\A=\{x\in\C  \colon \Pi(x)=\PiS,\, x\neq 0 \},
\end{equation}
which proves that it is non-empty and compact. We prove now the three items of the Lemma.

Item~(\ref{lem:V1}). Since $V(x)$
in (\ref{eq:newV}) is non-negative and zero if and only if $\Pi(x)=\PiS$, from expression (\ref{eq:setAequiv}) we obtain $V(x)=0$ if and only if $x\in\A$ and positive otherwise.
Moreover, as $x$ approaches zero, we have that $U_b(x)$ tends to zero, which implies $V(x) \to \infty$. Since $U_b(x) \leq \hat{U}$ for all $x$ due to the structure of $\C_0$, $\lim_{|x| \to \infty} \Pi(x) = +\infty$ implies $\lim_{|x| \to \infty} V(x) = +\infty$.

Item~(\ref{lem:V2}). This item follows in a straightforward way from the fact that $V(x)$ remains constant by construction during flow.

Item~(\ref{lem:V3}). First of all notice that $x=\!\smallmat{x_1\\x_2}\!\in\D$ implies $x_1=0$ and that $G(x)=\smallmat{\hat \theta \text{sign}(x_2) \\ x_2}$ for all $x\in \D\cap \Ba$.
We split the proof in three cases. We only consider jumps from points in the negative part of $\D$ (namely $x_2<0$) because of the central symmetry of the phase portrait. We also use the simplified notation $\Pi^+$ to denote 
$\Pi(x^+)$. Similar simplifications will be used for other quantities.

\begin{figure}[htbp]
\centering
{\includegraphics[width=0.49\columnwidth]{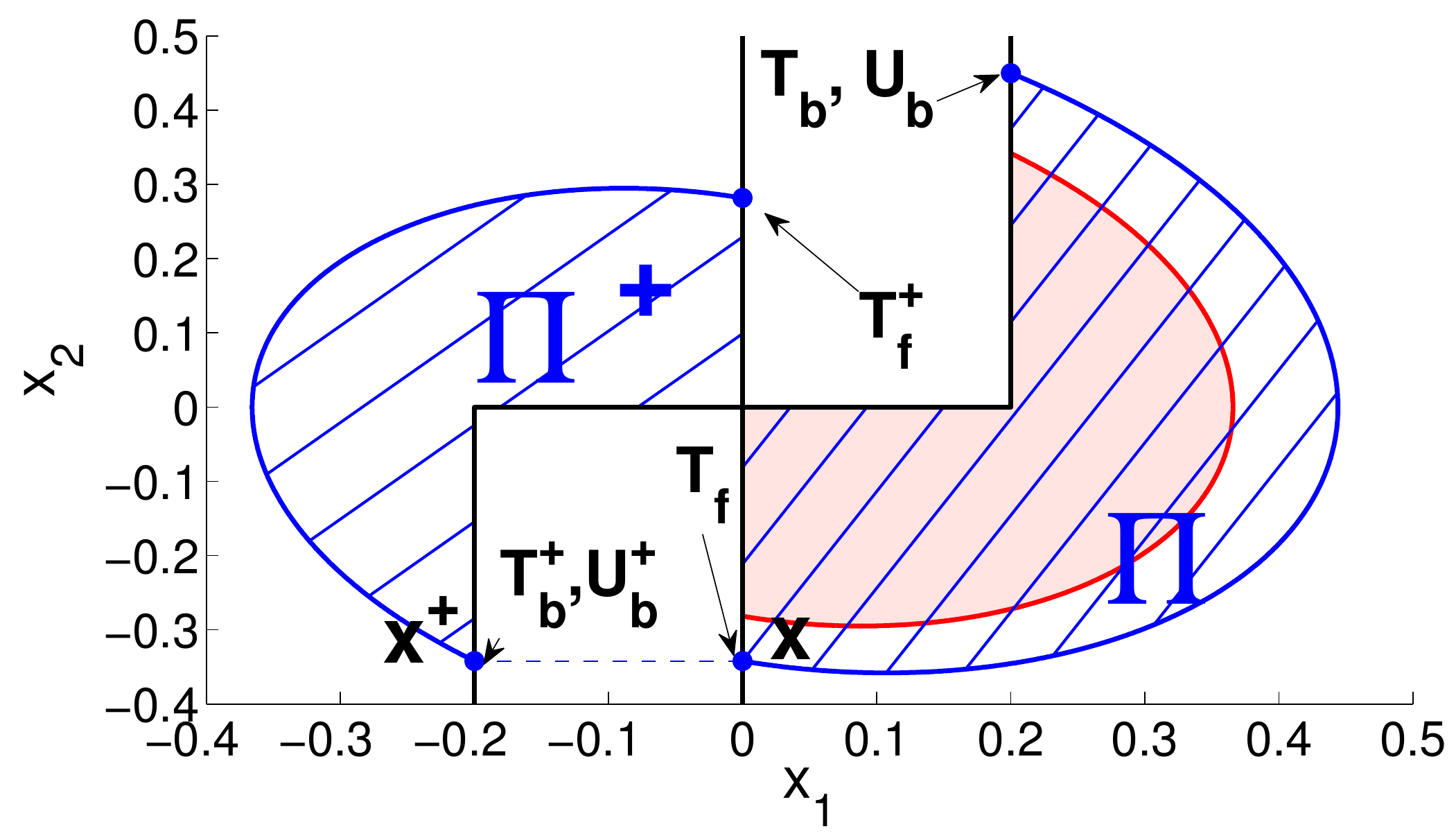}}
{\includegraphics[width=0.49\columnwidth]{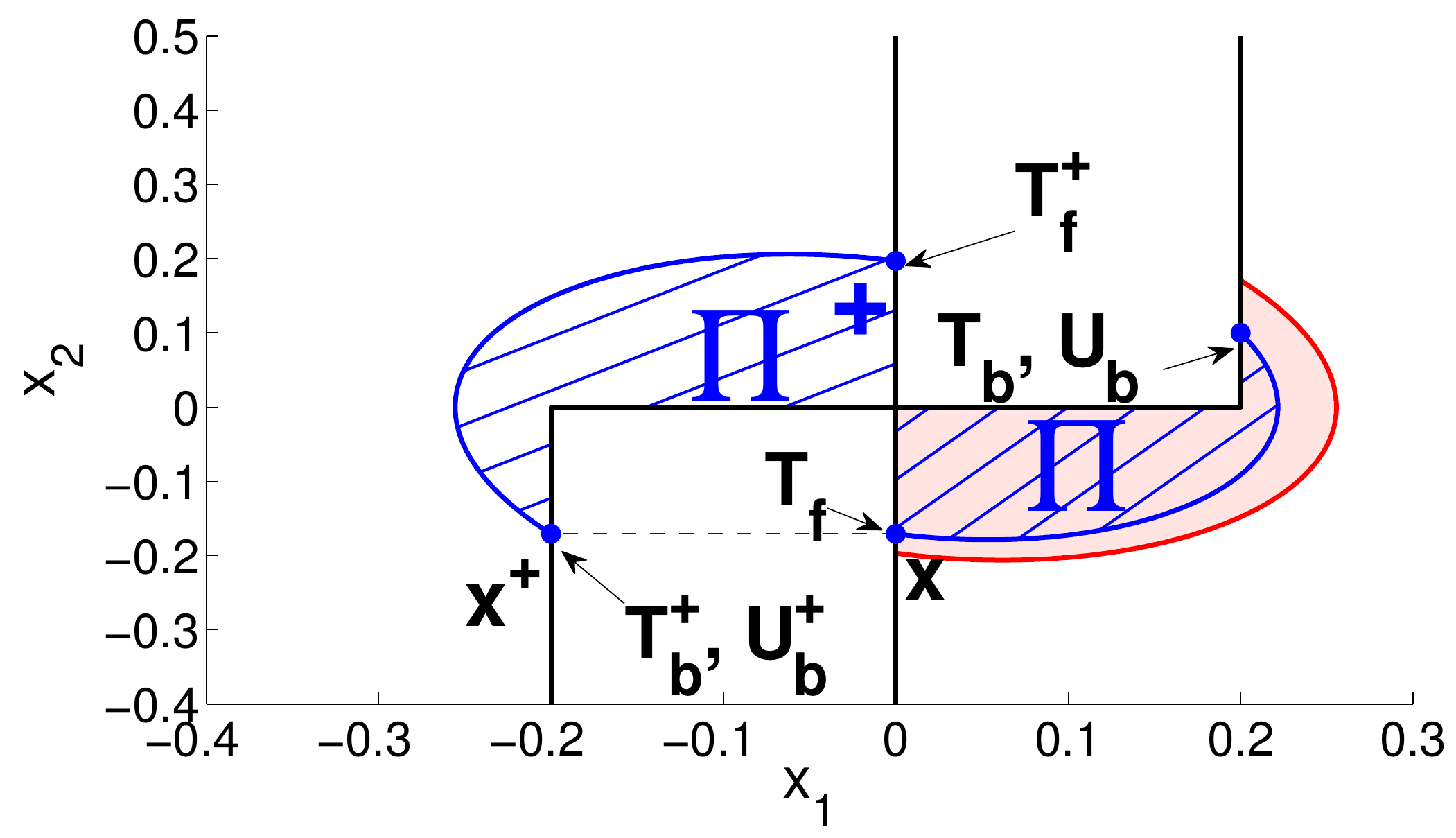}}
\caption{Left: $\Pi > \PiS > \Pi_0$. Right: $\PiS > \Pi > \Pi_0$. Pink areas are $\Pi^+$ mirrored about the origin.}
\label{fig:above_pi0}
\end{figure}

\emph{Case 1: $\Pi > \PiS > \Pi_0$}. 
First of all, by uniqueness of solutions $\Pi^+ > \PiS$, otherwise the flow would intersect the hybrid periodic orbit. 
Consider the left part of 
Figure~\ref{fig:above_pi0} and note that $\Pi > \PiS$ implies $\Pi^+ < \Pi$. 
Indeed, exploiting $U_b^+ = U_b = \hat{U} = c\Pi^\star$ and
$T_b + U_b = T_f + c\Pi$ and $T_b^+ = T_f $,
we get
$T_b = T_f + c\Pi - c\Pi^\star >  T_f = T_b^+$.
$\Pi > \Pi^+$ follows by monotonicity. 
Finally, the result is proven from
$$
0 = c \PiS - \hat{U} < c \Pi^+ - \hat{U} < c\Pi - \hat{U}.
$$

\emph{Case 2: $\PiS > \Pi > \Pi_0$}. 
First of all, $\Pi^+ < \PiS$ again from uniqueness of solutions. 
Consider the right-part of 
Figure~\ref{fig:above_pi0} and note that $\Pi_0 < \Pi < \PiS$ implies $\Pi^+ > \Pi$. 
In fact, following the argument of Case 1, 
we have that 
$U_b^+ = U_b = \hat{U} = c\Pi^\star$ and
$T_b + U_b = T_f + c\Pi$ and $T_b^+ = T_f $
thus 
$T_b = T_f + c\Pi - c\Pi^\star <  T_f = T_b^+$.
The result is proven from
$$
0 = c \PiS - \hat{U} > c \Pi^+ - \hat{U} > c \Pi - \hat{U}.
$$

\begin{figure}[htbp]
\centering
{\includegraphics[width=0.6\columnwidth]{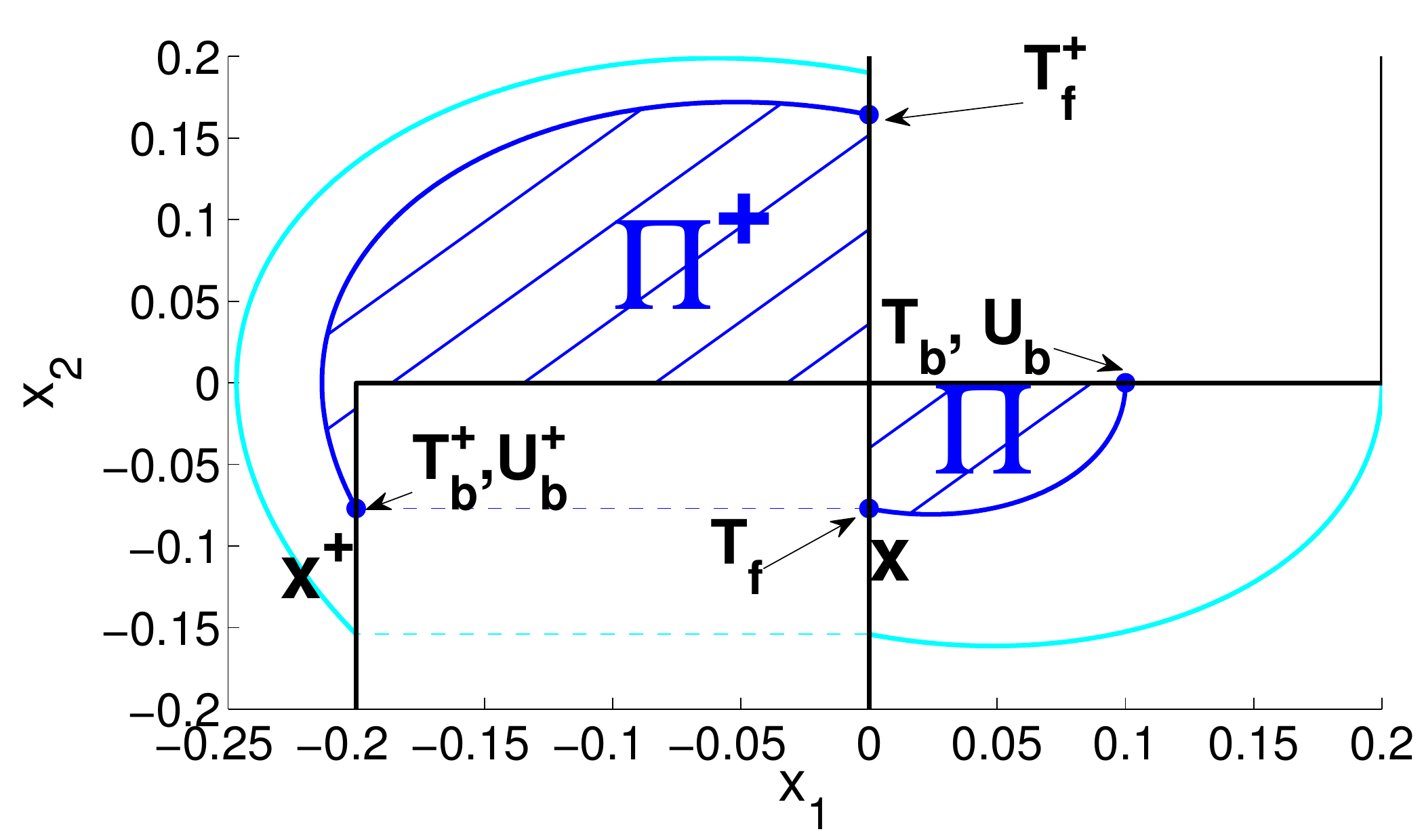}}
\caption{Case 3: $0<\Pi < \Pi_0$.}
\label{fig:in0s0}
\end{figure}

\emph{Case 3: $0<\Pi < \Pi_0$}. Consider Figure~\ref{fig:in0s0}.
In this case we have $U_b < \hat{U}$ because $U_b$ is evaluated on the horizontal part of $\C_0$. Then
\begin{align*}
V^+ - V & = {(c \Pi^+ - \hat{U})^2}\big/{\hat{U}} - {(c \Pi - \hat{U})^2 } \big/{U_b} \\
& < {(c \Pi^+ - \hat{U})^2}\big/{\hat{U}} - {(c \Pi - \hat{U})^2 }\big/{\hat{U}}.
\end{align*}
Now, observe that $c\Pi^+ < \hat{U} = c \Pi^\star$ because otherwise the forward solution from $x^+$ would intersect the flowing portion of the hybrid periodic orbit (thus contradicting uniqueness).
Then, using $U_b>0$, we get $V^+ -V<0$ from
\begin{align*}
&(c \Pi^+ - \hat{U})^2- (c \Pi - \hat{U})^2 \\
&\qquad = c(\underbrace{\Pi^+- \Pi}_{>0}) (\underbrace{c\Pi^+ -\hat{U}}_{<0} + \underbrace{c \Pi - \hat{U}}_{<0})< 0,
\end{align*}
where we used (i) $\Pi^+ > \Pi$ from $\Pi^+$ being evaluated from the vertical part of $\C_0$ and (ii) $c \Pi < \hat{U}$ from $c\Pi < U_b$.

\bibliographystyle{IEEEtranS}
\bibliography{IEEEabrv,refs}

% Generated by IEEEtranS.bst, version: 1.13 (2008/09/30)
\begin{thebibliography}{10}
\providecommand{\url}[1]{#1}
\csname url@samestyle\endcsname
\providecommand{\newblock}{\relax}
\providecommand{\bibinfo}[2]{#2}
\providecommand{\BIBentrySTDinterwordspacing}{\spaceskip=0pt\relax}
\providecommand{\BIBentryALTinterwordstretchfactor}{4}
\providecommand{\BIBentryALTinterwordspacing}{\spaceskip=\fontdimen2\font plus
\BIBentryALTinterwordstretchfactor\fontdimen3\font minus
  \fontdimen4\font\relax}
\providecommand{\BIBforeignlanguage}[2]{{%
\expandafter\ifx\csname l@#1\endcsname\relax
\typeout{** WARNING: IEEEtranS.bst: No hyphenation pattern has been}%
\typeout{** loaded for the language `#1'. Using the pattern for}%
\typeout{** the default language instead.}%
\else
\language=\csname l@#1\endcsname
\fi
#2}}
\providecommand{\BIBdecl}{\relax}
\BIBdecl

\bibitem{goebel2012hybrid}
R.~Goebel, R.~G. Sanfelice, and A.~R. Teel, \emph{Hybrid Dynamical Systems:
  modeling, stability, and robustness}.\hskip 1em plus 0.5em minus 0.4em\relax
  Princeton University Press, 2012.

\bibitem{Goebel2006}
R.~Goebel and A.~R. Teel, ``Solutions to hybrid inclusions via set and
  graphical convergence with stability theory applications,''
  \emph{Automatica}, vol.~42, no.~4, pp. 573--587, 2006.

\bibitem{Grasman1987}
J.~Grasman, \emph{Asymptotic methods for relaxation oscillations and
  applications}, 1st~ed., ser. Applied Mathematical Sciences.\hskip 1em plus
  0.5em minus 0.4em\relax Springer, 1987.

\bibitem{Gunther2013}
F.~Gunther and F.~Iida, ``Preloaded hopping with linear multi-modal
  actuation,'' in \emph{IEEE/RSJ International Conference on Intelligent Robots
  and Systems}, 2013, pp. 5847--5852.

\bibitem{Hirsch1974}
M.~Hirsch and S.~Smale, \emph{Differential Equations, Dynamical Systems, and
  Linear Algebra (Pure and Applied Mathematics, Vol. 60)}.\hskip 1em plus 0.5em
  minus 0.4em\relax Academic Press, 1974.

\bibitem{lakatos2014switching}
D.~Lakatos and A.~Albu-Sch{\"a}ffer, ``Switching based limit cycle control for
  compliantly actuated second-order systems,'' in \emph{Proceedings of the 19th
  IFAC World Congress}, 2014, pp. 6392--6399.

\bibitem{lakatos2014jumping}
D.~Lakatos, G.~Garofalo, A.~Dietrich, and A.~Albu-Sch{\"a}ffer, ``Jumping
  control for compliantly actuated multilegged robots,'' in \emph{IEEE
  International Conference on Robotics and Automation}, 2014, pp. 4562--4568.

\bibitem{lakatos2013modal}
D.~Lakatos, G.~Garofalo, F.~Petit, C.~Ott, and A.~Albu-Sch{\"a}ffer, ``Modal
  limit cycle control for variable stiffness actuated robots,'' in \emph{IEEE
  International Conference on Robotics and Automation}, 2013, pp. 4934--4941.

\bibitem{lakatos2013nonlinear}
D.~Lakatos, F.~Petit, and A.~Albu-Sch{\"a}ffer, ``Nonlinear oscillations for
  cyclic movements in variable impedance actuated robotic arms,'' in \emph{IEEE
  International Conference on Robotics and Automation}, 2013, pp. 508--515.

\bibitem{meirovitchVibrations}
L.~Meirovitch, \emph{Fundamentals of Vibrations}.\hskip 1em plus 0.5em minus
  0.4em\relax McGraw-Hill Companies, 2000.

\bibitem{PrieurTAC14}
C.~Prieur, A.~Teel, and L.~Zaccarian, ``Relaxed persistent flow/jump conditions
  for uniform global asymptotic stability,'' \emph{IEEE Transactions on
  Automatic Control}, vol.~59, no.~10, pp. 2766--2771, 2014.

\bibitem{saglamlyapunov}
C.~O. Saglam, A.~R. Teel, and K.~Byl, ``Lyapunov-based versus {P}oincar{\'e}
  map analysis of the rimless wheel.'' in \emph{IEEE 53rd Conference on
  Decision and Control}, 2014, pp. 1514--1520.

\bibitem{teel2013stabilization}
A.~R. Teel, R.~Goebel, B.~Morris, A.~D. Ames, and J.~W. Grizzle, ``A
  stabilization result with application to bipedal locomotion,'' in \emph{IEEE
  52nd Conference on Decision and Control}, 2013, pp. 2030--2035.

\bibitem{Westervelt2007}
E.~Westervelt, J.~Grizzle, C.~Chevallereau, J.~Choi, and B.~Morris,
  \emph{Feedback control of dynamic bipedal robot locomotion}, ser. Control and
  automation.\hskip 1em plus 0.5em minus 0.4em\relax Boca Raton: CRC Press,
  2007.

\end{thebibliography}

\end{document}